\documentclass[runningheads]{llncs}
\usepackage[T1]{fontenc}
\usepackage{amsmath}
\usepackage{graphicx}
\usepackage{slashbox}
\usepackage{amsmath}
\begin{document}
\title{An Identity Based Agent Model for AI Value Alignment}
%
%
\author{
Karthik Sama\inst{1,2}\orcidID{0000-0002-5538-916X} \and
Janvi Chhabra\inst{1}\orcidID{0000-0002-4642-7726} \and
Arpitha Srivathsa Malavalli\inst{1}\orcidID{0000-0002-5491-5535} \and
Jayati Deshmukh\inst{1,3}\orcidID{0000-0002-1144-2635} \and
Srinath Srinivasa\inst{1,}\orcidID{0000-0001-9588-6550}
}
\authorrunning{K. Sama et al.}
%
\institute{
International Institute of Information Technology Bangalore, Bengaluru, India\\
\and
KU Leuven, Leuven, Belgium;
\email{saikarthik.sama@student.kuleuven.be}
\\
\and
University of Southampton, Southampton, UK;
}
%
\maketitle              
\begin{abstract}
 With AI systems being deployed across diverse societal contexts, the AI alignment problem has become critical. Ensuring that AI systems act in accordance with human values, particularly in complex multi-agent environments, remains a significant challenge. Conventional approaches—typically employing uniform value representations and consequentialist methods—often fail to capture the inherent variability in how values influence individual agents' decision-making. In this work, we address this gap by extending the Computational Transcendence (CT) framework to integrate an agent's sense of self into its decision-making process. Our approach embeds values within an agent’s identity, thereby facilitating adaptive, individualized behavior that accounts for external social influences, such as conformity. We demonstrate our model in a multi-agent simulation of transit decision-making, where agents choose between public and private transportation based on their value-driven identities. Our results suggest that using identity as a basis for value alignment offers a promising pathway for capturing human-like decision-making in AI systems.
\end{abstract}
\section{Introduction}

%


As AI systems become increasingly integrated into various aspects of society, ensuring their alignment with human values and ethical principles of society has become a central concern. This challenge, known as the AI alignment problem, encompasses multiple facets. In applications such as recommendation systems, objectives like transparency and robustness may suffice to ensure appropriate behavior. However, more complex domains like healthcare decision support, judicial advices and autonomous vehicles, demand nuanced approaches to alignment. This complexity is inherent multi-agent systems, where both human-AI and AI-AI interactions might be possible, necessitating a deeper understanding of the alignment problem. Addressing these challenges involves disciplines such as Machine Ethics, which focuses on integrating ethical reasoning capabilities into AI systems, and AI Value Alignment, which aims to ensure that artificial agents' decisions are congruent with human values. In this work, we contribute to the AI Value Alignment problem by drawing on an existing framework of identity based decision-making to systematically incorporate values into the decision-making process of agents.

\subsection{AI Value Alignment}


In human decision making values play a significant role. The work by Kluckhohn~\cite{Kluckhohn_value_action} defines values as "conceptions of the desirable, guiding individuals' selections", emphasizing their role in directing actions toward desired goals. Furthermore, these values serve as ethical compasses at the individual level~\cite{fritzsche1991model}, thereby reinforcing the need to incorporate values as a foundational element of decision-making in artificial systems.

This observation gives rise to the problem of Value Alignment in AI, where the goal is to construct formal models of agents that can internalize values in a manner consistent with the human contexts in which they operate. The value alignment problem can be subdivided into two key components~\cite{ai_values_alignment}: (i) representing or encoding values, and (ii) transforming encoded values into autonomous decision-making processes. 

Encoding abstract values is inherently challenging. As highlighted by Schwartz's theory~\cite{schwartz2012}, values fundamental to human nature can be both conflicting, and different combinations of values may lead to similar behaviors. The problem of constructing value representations has been approached from multiple perspectives. In Cooperative Inverse Reinforcement Learning~\cite{hadfield2016} artificial agents interact with humans to reconstruct a reward function assuming human actions are optimally guided by values. Nardine et al.~\cite{frameworkHumanValues} decompose values into directed acyclic graphs (DAGs) -- with an abstract value forms the root, with each node breaking down into simpler child nodes, and the leaf nodes representing measurable properties in the context of interest.

In this work, we focus on the challenge of translating encoded value representations into decision-making. This problem has been primarily tackled by using consequentialist approaches such as aggregating preferences over possible outcomes~\cite{value_alignment_formal}, designing external norms that maximize value alignment~\cite{montes2022synthesis} and employing the reward functions learned from Inverse Reinforcement Learning approaches as the policy for RL agents~\cite{hadfield2016}. However, previous research has shown that individuals have inherent variability in how values influence their decisions~\cite{schwartz2012}. Consequently, in multi-agent settings, existing methods often struggle to scale or to capture this inherent variability when a uniform value representation is assumed for an entire society. To address this, we shift our focus to agent models grounded in identity or a sense of self, wherein the variability of value associations is incorporated as an integral aspect of each agent's identity.

\subsection{Identity based Decision making frameworks}
Modeling autonomy in artificial agents has been explored through various paradigmatic standpoints~\cite{Paradigms_of_Computational_Agency}. Among these, the Models of Self paradigm conceptualizes agents with an internal identity which forms the basis of their decision making. Agent-based modeling grounded in the notion of self has been applied across diverse contexts—for instance, computational sense of control enhancing autonomy of motor control in unpredictable environments~\cite{kahl2022}, enhancing social interaction through integrating self-awareness in decision-making~\cite{subagdja2021}, fostering cooperation in non-cooperative games by modeling agents who identify with each other~\cite{deshmukh2022computational}.

One such identity-based approach is the Computational Transcendence (CT) framework~\cite{deshmukh2022computational}. It is relevant for the problem of translating encoded values since enables modeling agent's identity as the basis for it's decision-making. An agent's identity set may include other agents, group affiliations (e.g., communal societies, nationalities), or abstract concepts like human values. While the CT framework is well-suited for capturing identity-driven behavior, it assumes the ability to compute utilities associated with identity objects -— an assumption that becomes problematic for abstract values, whose impact on decisions is not directly measurable. In this work, we address this limitation by extending the CT framework to map value representations to quantifiable observables. Our choice of the CT framework is motivated by its ability to -- (i) Capture how values variably influence individual agent's decisions. (ii) Facilitate modeling of adaptive agents, making the framework generalizable. (iii) Enable integration of external social influences and incentives, facilitated by its consequentialist structure.


We extend the Computational Transcendence (CT) framework to develop a multi-agent simulation for modeling decision-making in the context of transit behavior. In this setting, agents choose between public (bus) and private (taxi) modes of transportation. Each choice results in specific outcomes—such as time, cost, carbon footprint, and comfort—which we refer to as measurable observables. Agents’ decisions are guided by abstract values embedded in agent's identity which include frugalism, idealism, individualism, and pragmatism.

Crucially, the extended framework enables agents to adapt their identity associations over multiple trips based on experiential feedback and contextual dynamics. We examine how agents’ decisions are shaped not only by internalized value incorporate but also by external social factors, particularly conformity to study their dynamic influence on decision making. This allows us to simulate more human-like decision-making processes, capturing both the how different values variably influences decisions, and inherent variability in how individuals associate with these values.

\section{Modelling Agents that Identify with Values}\label{sec:model}

\subsection{Computational Transcendence}
The Computational Transcendence(CT) model~\cite{deshmukh2022computational} provides a framework to build agents based on identity, allowing an innate sense of self based on internal associations to drive decision making. An agent $a$ in CT is defined with an elastic sense of self represented by $S(a) = (I_a, d_a, \gamma_a)$ where: 

\begin{itemize}
    \item $I_a$ represents the set of objects with which an agent $a$ identifies itself.
    \item $d_a: \{a\} \times I_a \rightarrow \Re^+$ is a set of semantic distances corresponding to each identity object, where these distances represent how close or important are specific identity objects to an agent.
    \item $\gamma_a \in [0,1]$ represents the transcendence level of the agent $a$. It captures agent's capability of associating its identity with different objects.
\end{itemize}

A simple way to think about elastic sense of self if through a person's friends circle. Each friend represents an identity object, degree of friendship being captured by semantic distances, while transcendence level influencing the size of friends circle. The utility of a choice $c$ an agent makes is then computed as weighted utility incurred by different identity objects, where the weights $\gamma_d^{d_a(o)}$ are attenuation factors~\cite{deshmukh2022computational}. This is formalized as follows,
\begin{equation}\label{eqn:orgUtil}
    u_{c}(a) = \frac{\sum_{j=1}^{m}\gamma_a^{d_a(o_j)}.u_{c}(o_j)}{\sum_{j=1}^{m}\gamma_a^{d_a(o_j)}}
\end{equation}

Using these perceived utilities, agent $a$ makes decisions. However, this formulation implicitly assumes that $u_{c}(o)$, the utility of a choice $c$ with respect to an identity object $o$, is computable for all $o$. In the case of abstract values, this assumption is problematic, as determining the utility of a choice based on its alignment with an abstract value is challenging.

We notice that the identity objects and the measurable quantities in a context, whose utilities are calculable needn't be the same. We refer to these measurable quantities as contextual observables, in the pretext of value representation work by Nardine et. al. \cite{frameworkHumanValues} these entities are similar to the verifiable properties in the system. The objects in the identity set of an agent are invariant across different contexts, while the contextual observables can change. Consider an example where an agent identifies with an environmentalism. The contextual observables corresponding to environmentalism can change as the context changes -- in the context of making transit choice the relevant observable for environmentalism could be carbon footprint while in a different context of choosing a drink, eco-friendly packaging can be seen as the observable. Addressing the gap between identity objects and contextual observables would enable modeling decision making in agent based on abstract values using the CT framework.

\subsection{Schemas to Bridge Identity Objects with Observables}

To mathematically model how identity objects of an agent interact with the measurable observables of a context, we introduce schemas. A schema $\mathbf{S}$ is a row-stochastic rectangular matrix defined for in a particular context, with it's rows corresponding to identity objects, and columns corresponding to contextual observables. Each entry  $\mathbf{S_{ij}}$ of the schema matrix corresponds to the weight given by the $i^{th}$ identity object to $j^{th}$ contextual observable. A weight $0$ means the a specific observable is irrelevant to an identity object. While a weight of $1$ implies identity object and contextual observable are identical. If all the identity objects have corresponding contextual observables then $\mathbf{S}$ becomes an Identity matrix $\mathbf{I}$. The row-stochasticity of $\mathbf{S}$, where each row of $\mathbf{S}$ sums up to 1, ensures that each identity object has a normalized distribution over the contextual observables. 

In Section~\ref{sec:model}, we model transit behavior using a multi-agent framework where Table~\ref{tab:schemavalue} provides an example schema of the abstract values over the measurable observables relevant for transit choices. In the context of value alignment, these schemas are representations of values in terms of a vector of measurable properties. Thereby, schemas act as a bridge for calculating utilities of encoded value representations enabling the use of CT framework to simulate agent decision-making. It is important to emphasize that the design of these schemas must be carefully guided by domain experts to ensure the effectiveness and realism of agent-based simulations. However, more data-driven approaches\cite{frameworkHumanValues} to represent values in terms of measurable properties are also being actively developed and can be used to device schemas. We next present our extended CT framework based on schemas.

\subsection{Extended CT framework}

We now present our extended CT framework. Consider the identity set of $a$ to consist of $m$ identity objects -- $I_a: \{o_1, o_2, ..., o_m\}$. The attenuation factors $\gamma_a^{d_a(o)}$ which represent how important an identity object is to an agent follow from the original CT framework\cite{deshmukh2022computational}. In a defined context, let there be $n$ contextual observables. The utility $a$ derives by making a choice can be written as follows,

\begin{equation}\label{eqn:Extended_CT}
    u_c(a) = \frac{1}{\Sigma_{i=1}^{m}\gamma^{d_a(o_i)}}
    \begin{bmatrix}
        \gamma^{d_a(o_1)},\text{ .. },
        \gamma^{d_a(o_m)}
    \end{bmatrix}_{1 \times m} \mathbf{S_{m \times n}}
    \begin{bmatrix}
        u(co_1), \text{ .. }, u(co_n)
    \end{bmatrix}^T_{1 \times n}
\end{equation}

The utility calculation of a choice in Equation~\ref{eqn:Extended_CT} can be understood as follows,
\begin{itemize}
    \item [\textbullet] The normalized attenuation factors, represented as $(1 \times m)$ row vector, depict how agents sense of self is decomposed into different identity objects.
    \item [\textbullet] Each row of schema $\mathbf{S_{m \times n}}$ further decomposes these identity objects in terms of measurable observables in a given context.
    \item [\textbullet] Finally we have a $(n \times 1)$ column vector of utilities of observables incurred by taking choice in the defined context.
\end{itemize}

When identity objects and measurable quantities are identical, $\mathbf{S_{m \times n}}$ becomes Identity matrix $(\mathbf{I_{n}})$, resulting in Equation~\ref{eqn:Extended_CT} simplifying to Equation~\ref{eqn:orgUtil} as described in original CT framework~\cite{deshmukh2022computational}. We can use the computed utility of each option to probabilistically model an agent’s decision-making during a simulated run.

\subsection{Adaptive Agents via Identity Association Updates}\label{subsec:dist_update}
The CT framework enables the modeling of adaptive agents that adjust their semantic distances based on the accumulated utilities of their decisions. More specifically, agents compute the marginal utility contributed by an identity object over an epoch consisting of multiple decision-making rounds~\cite{deshmukh2022computational}. At the end of each epoch, if the marginal utility exceeds a predefined threshold, the agent strengthens the association by decreasing its semantic distance to the corresponding identity object. Conversely, if the utility falls below the threshold, the semantic distance is increased.

In the extended CT framework, defining the marginal utility for each identity object enables the direct application of the adaptive mechanism introduced in the original model. Given that the schema matrix $\mathbf{S}$ distinguishes each identity object on a per-row basis, this computation becomes straightforward. The marginal utility associated with an identity object $o_i$ over an epoch comprising $R$ decision-making runs is computed as follows,

\begin{equation}\label{marginal_util}
epoch\_utility = \Sigma_{r=1}^{R} (\mathbf{S_{i,:}} \cdot [u_r(co_1), .. , u_r(co_n)]^T)    
\end{equation}

In Equation~\ref{marginal_util}, $\mathbf{S}_{i,:}$ denotes the $i^{\text{th}}$ row of the schema matrix $\mathbf{S}$, and $u_r(co)$ represents the utility of a contextual object in the $r^{\text{th}}$ decision-making run. The dot product between an identity object's schema and the contextual utilities yields the marginal utility associated with that identity in a given run. This value is aggregated over an epoch to update semantic distances. In the extended CT framework we allow utilities to be negative for observables and do not distinguish between utilities and costs.

An agent is considered \textit{stabilized} when it no longer updates its semantic distance to any identity object—indicating that it has no further incentive to adapt. In our experiments, we analyze the state of a population of such stabilized agents.

\section{Modeling Transit Choices Using the Extended CT Framework}\label{sec:transitChoices}

In this section, we use our theoretical extension of the CT framework to model multi-agent behavior in transit decision-making, where agents choose between taxi and bus—representing private and public transportation, respectively.

\subsection{Contextual Observables underlying Transit Choices}
In the context of transit choices, we consider four contextual observables: cost, time, congestion, and carbon footprint. For our experiments, we selected two locations within a metropolitan area (Bangalore, India) and used Google Maps to estimate the mean and variance of travel time and cost for both transit modes. We assume zero wait times for vehicle availability, and carbon footprint is computed as the product of total distance traveled and average carbon emission per vehicle. Below we describe these contextual observables and their modeling in detail:

\begin{description}
    \item[Cost:] The monetary cost incurred by an agent for the chosen transit choice. A constant function is used to model the cost of transit for a given vehicle. For transit between the two selected points, the average cost of a bus is 20 units, and for a taxi is 300 units. 
    \item[Time:] The time an agent takes from source to destination in the chosen transit choice. A right-skewed Gumbel distribution\cite{railTransitGumbel2016} is used to model vehicle travel times, as it effectively captures rare, unforeseen events that can cause unusually high delays like accidents or traffic jams. For taxis, this distribution is instantiated with a mean of 20 minutes and a variance of 5 minutes. For buses, the respective values are 47 minutes and 10 minutes.
    \item[Congestion:] In the transit choice made by agent congestion is given by,
    \begin{equation}\label{eqn:congestion}
        \text{Congestion} = \frac{\text{Occupancy of Vehicle}}{\text{Seating Capacity}}    
    \end{equation}
      Each vehicle(transit choice) has a fixed \textit{seating capacity} and \textit{maximum occupancy}: for taxis, these are 4 and 5, respectively; for buses, 40 and 80. Vehicle occupancy in a trip is modeled probabilistically. Taxi occupancy is represented by a discrete distribution over values 1–5: \{1{:}~0.1,\ 2{:}~0.2,\ 3{:}~0.3,\ 4{:}~0.3,\ 5{:}~0.1\}, while bus occupancy follows a Gaussian distribution with mean 40 and variance 25.
     \item[Carbon footprint:] In the transit choice made by agent carbon footprint is given by,
    \begin{equation}
        \text{Carbon footprint} = \frac{\text{Total Carbon Emission per trip}}{\text{Occupancy of Vehicle}}  
    \end{equation}
     Taxis emit an average of 40 grams of carbon per kilometer, while buses emit 200 grams per kilometer. \textit{Occupancy} is sampled as previously described.
\end{description}

\subsection{Perceived Utilities of Observables}
The observable quantities differ in scale and units, and therefore must be transformed before being incorporated into Equation~\ref{eqn:Extended_CT}. Behavioral economics offers principled methods to transform measurable observables into perceived utilities, allowing us to better model agent decision-making. We apply the following transformations:

    \begin{itemize} \item[\textbullet] \textbf{Prospect Theory}~\cite{kai1979prospect} is used to model the perceived utility of time, cost, and carbon footprint. This transformation curve that resembles an asymmetric sigmoid: steeper for losses than gains (capturing loss aversion).
    \item[\textbullet] \textbf{Congestion} is transformed using a shifted ReLU function: $2 - \max(1, \text{congestion})$. As defined in Equation~\ref{eqn:congestion}, when congestion is less than or equal to 1, every passenger is seated, resulting in a utility of 1. When congestion exceeds 1, the perceived utility decreases linearly.
\end{itemize}

\subsection{Value Considerations in Transit Choices}\label{subsubsec:transitValues}

In this simulation of transit choices, we hypothesize the following values driving multi-agent behavior:

\begin{description}
    \item[Frugalism:] The quality of being resource-conscious, particularly with regard to money. In this context, it favors choices that minimize monetary cost.
    \item[Idealism:] A value orientation concerned with achieving utopian or morally ideal goals. Here, it promotes environmentally sustainable behavior, favoring choices with minimum carbon footprint.
    \item[Individualism:]  A focus on personal autonomy and individual identity. In this context, it emphasizes personal comfort and incurring less travel time for the commute.
    \item[Pragmatism:] A practical, outcome-oriented value that emphasizes efficiency. In this context, it considers all practical aspects—such as time, cost, and comfort of a transit choice.
\end{description}

Table~\ref{tab:schemavalue} encapsulates the discussed influence of values on observable quantities involved in transit through a schema.

\begin{table}[ht]
    \vspace{-1em}
    \renewcommand{\arraystretch}{1.5}
    \begin{center}
        \begin{tabular}{|c|c|c|c|c|}
            \hline
            {\backslashbox{Value}{Observable}} & Cost & Time & Congestion & C Footprint\\
            \hline
            Frugalism & $\frac{3}{5}$& $\frac{1}{5}$& $\frac{1}{5}$& 0\\
            \hline
            Idealism & $\frac{1}{10}$ & $\frac{1}{10}$&  $\frac{1}{10}$ & $\frac{7}{10}$\\
            \hline
            Individualism & $\frac{2}{10}$ & $\frac{3}{10}$ & $\frac{5}{10}$ & 0\\
            \hline
            Pragmatism & $\frac{1}{3}$ & $\frac{1}{3}$ & $\frac{1}{3}$ & 0\\
            \hline
        \end{tabular}
    \end{center}
    \caption{Schemas of values over observables}
\end{table}\label{tab:schemavalue}


\subsection{Incorporating Conformity as an External Social Factor}
While an agent's identity models internal factors influencing decision-making, external social factors arise from the network or society in which the agent is embedded. To enhance the realism of our multi-agent simulation, we incorporate conformity as an external social influence on agent behavior. The CT framework’s utility-based structure allows for seamless integration of socially derived utility. We model conformity through a network of agents, where each agent’s decisions are influenced by the transit choices of its neighbors. Let $\mathit{frac\_neigh}_{c_i}$ denote the fraction of neighbors selecting choice $c_i$. The utility of choice $c_i$ is then augmented as follows:

\begin{equation}\label{eqn:conform} u(c_i) += cf \cdot \mathit{frac\_neigh}_{c_i} \end{equation}

Here, $cf \in [0,1]$ represents the degree to which agents in a society conform to their peers.

\section{Experiments and Results}\label{sec:expandresults}
We run the transit choice simulation for agents embedded in a social network based on Erdős–Rényi graph with $500$ agents. Each agent is initialized with a transcendence level $\gamma = 0.8$ ensuring that each abstract value discussed in Section~\ref{subsubsec:transitValues} significantly impacts early decision making of agents. Each epoch consists of $10$ trips so that agents accumulated enough evidence to update their semantic distances. Our experiments examine which values most influence decisions, how initial societal configurations affect stabilized states, and the impact of conformity on population-level preferences.

\subsection{Impact of Values on populations' choices}\label{subsec:ImpactValues}

We begin with understanding how the identification with different values evolves for the population, and assume there is no conformity in the population while the initial semantic distances of the population are uniformly sampled -- $d \sim U[0,4]$.

\vspace{-5pt}
\begin{table}[ht]
\centering
\begin{tabular}{|c c|}
\hline
\multicolumn{2}{|c|}{Whole population} \\
\multicolumn{2}{|c|}{\includegraphics[width=0.60\columnwidth]{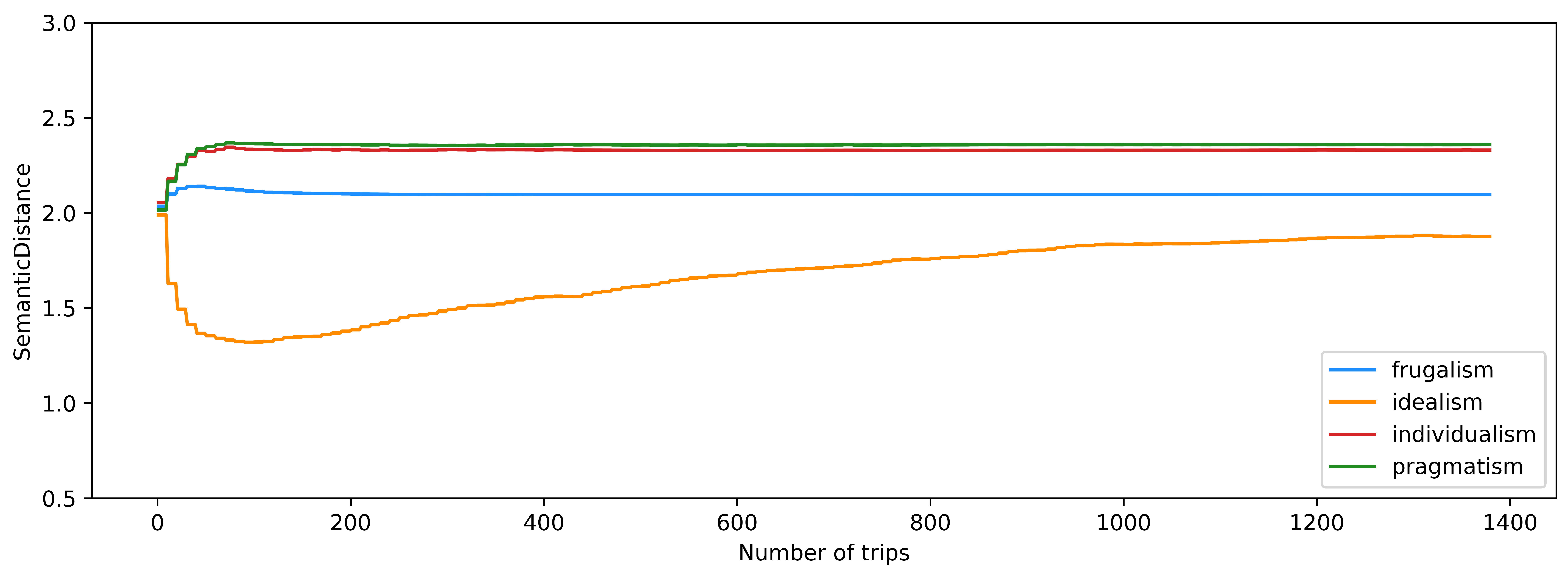}} \\
\hline
\multicolumn{2}{|c|}{Sub-populations} \\
{\includegraphics[width=0.5\columnwidth]{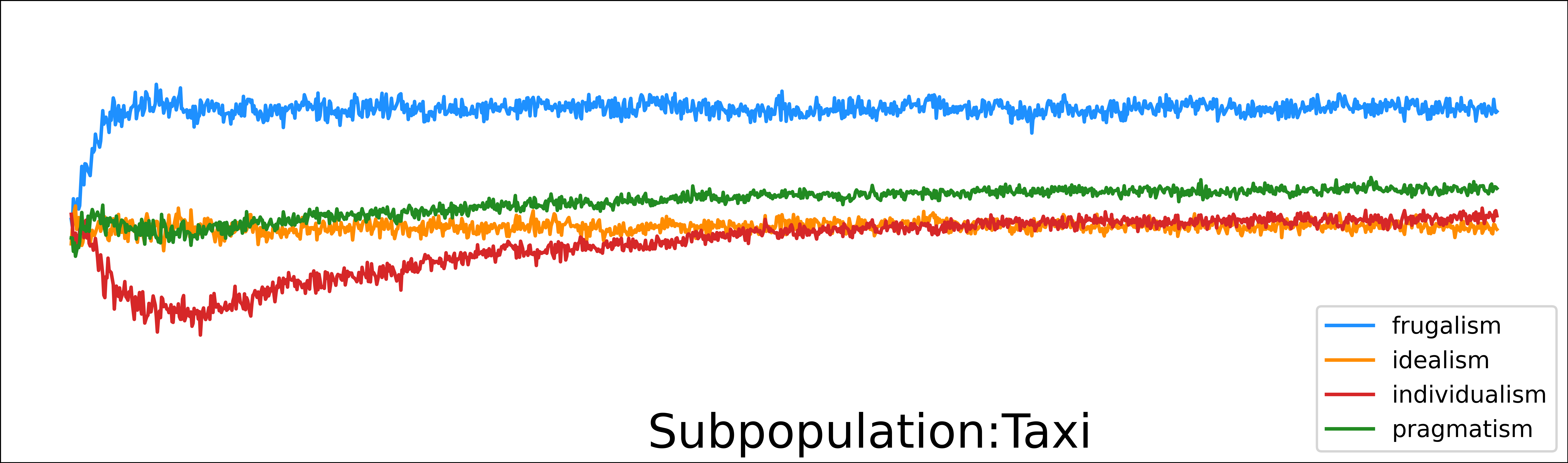}} & {\includegraphics[width=0.5\columnwidth]{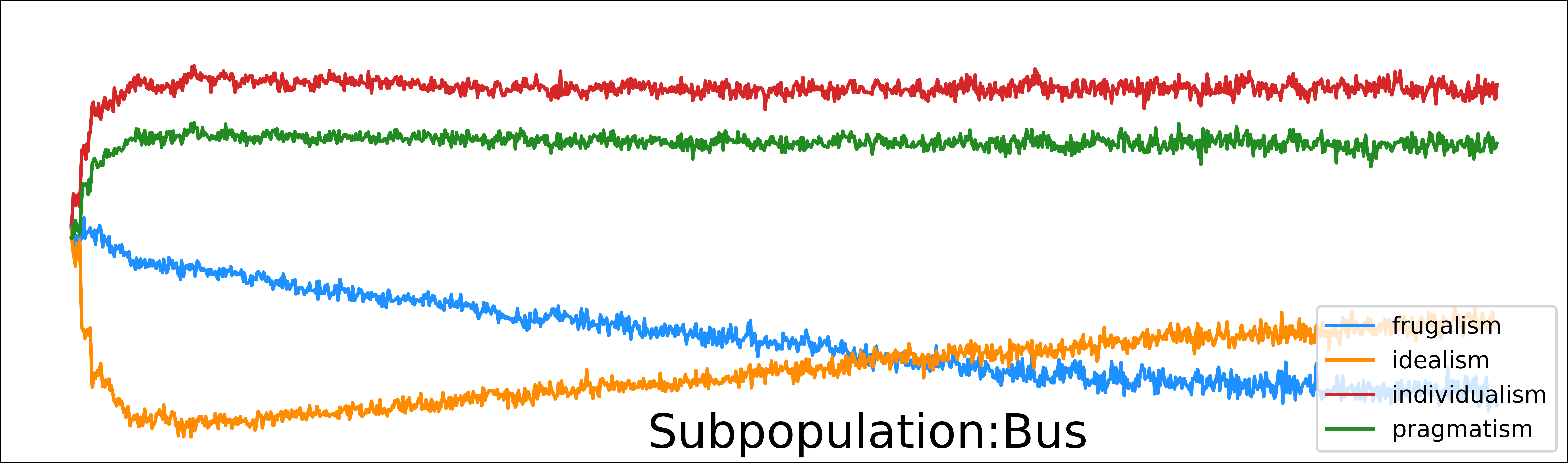}} \\
\hline
\end{tabular}
\caption{Trends of average semantic distances (identity associations) to different values across the trips for the sub-populations that chose taxi and bus. \textit{Note that the noise in these trends arises from the stochasticity in individual agent's decisions.}}\label{tab:subSemDist}
\end{table}
\vspace{-20pt}

The first row of Table~\ref{tab:subSemDist} shows how the average semantic distances, and consequently the population's identification with different values, evolve over the course of the simulation. Toward the end, these distances exhibit minimal change, indicating that the population has reached a steady state. Under this configuration, we observe that $66.5\%$ of the stabilized population opts for private transport. Interestingly, idealism emerges as the most prominent value, which appears to contradict the population’s preference for taxis—given that they emit more carbon per trip on average.

In the second row of Table~\ref{tab:subSemDist}, we examine the identity associations of sub-populations grouped by their transit choices. Idealism remains a prominent value across the population, since even within the taxi subpopulation its association declines only marginally—likely due to other contextual observables offsetting the negative utility from carbon emissions. We also observe the following from the sub-populations:

\begin{itemize}
    

    \item[\textbullet] Association to values like frugalism and individualism varies significantly between the subpopulations. Thus, they strongly influence the transit choices. 
    
    \item[\textbullet] In contrast, association to idealism and pragmatism varies slightly between these subpopulations. Suggesting that they weakly influence transit choices.
    
    
\end{itemize}

\subsection{Sensitivity of Initial Semantic Distances}\label{subsec:initialSD}

We next investigate how initial semantic distances influence the final state of the population. In each experiment, we systematically reduce the initial association of the population with a specific abstract value ($v_s$). To achieve this, we sample the initial semantic distance to $v_s$ from a uniform distribution $d_{v_s} \sim U[0,2]$, while the distances to all rest of the values ($v_r$) are sampled from $d_{v_r} \sim U[0,4]$. Table~\ref{tab:initialDistancesVary} summarizes the four resulting configurations. In the corresponding plots, blue and orange bars represent the average semantic distances to each value in the initial and final states, respectively.

\vspace{-5pt}
\begin{table}
    \centering
    \begin{tabular}{|c|c|}
        \hline
        Frugalism - Public transit: $64.2\%$ & Idealism - Public transit: $47.6\%$\\
       \includegraphics[width=0.4\textwidth]{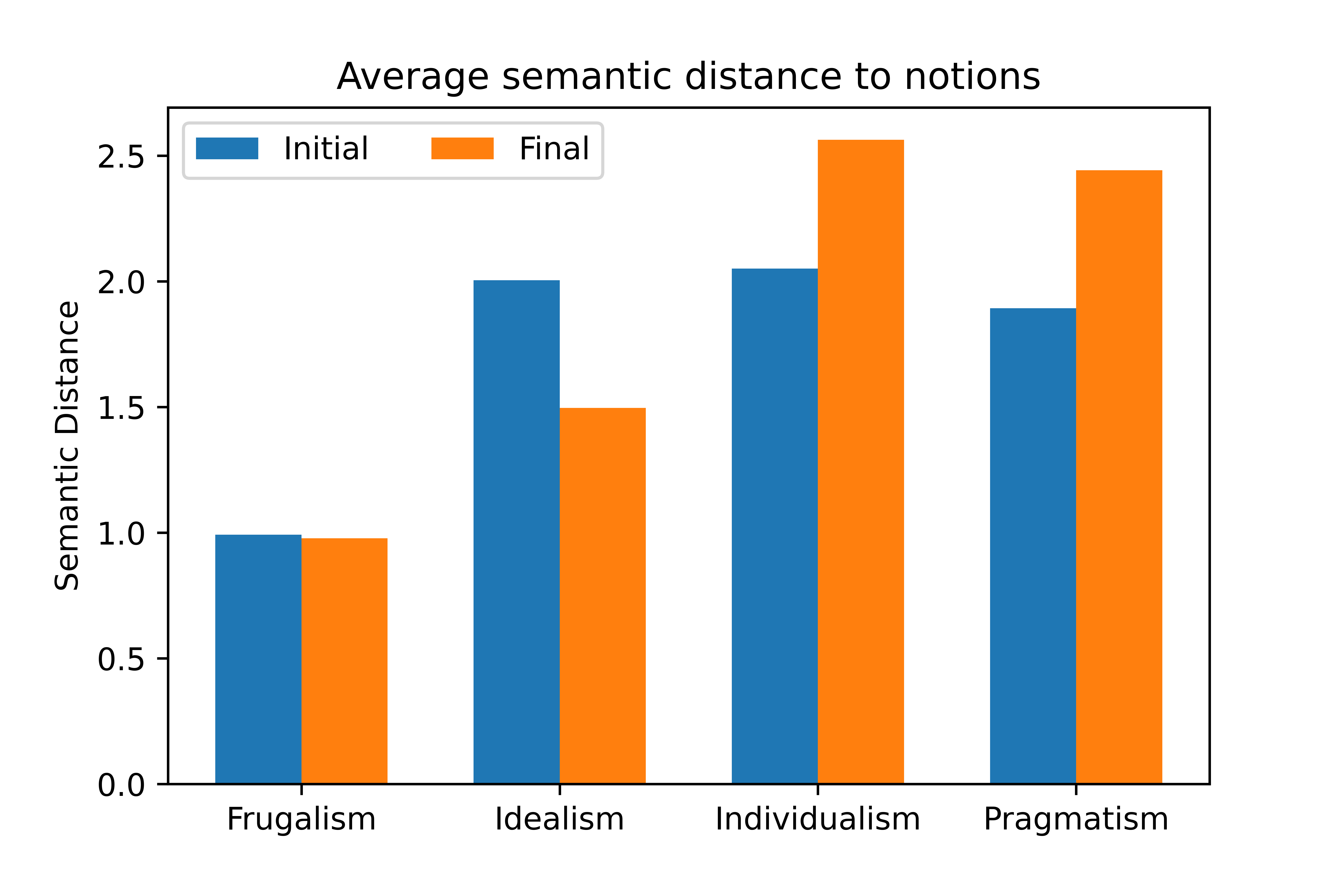} & \includegraphics[width=0.4\textwidth]{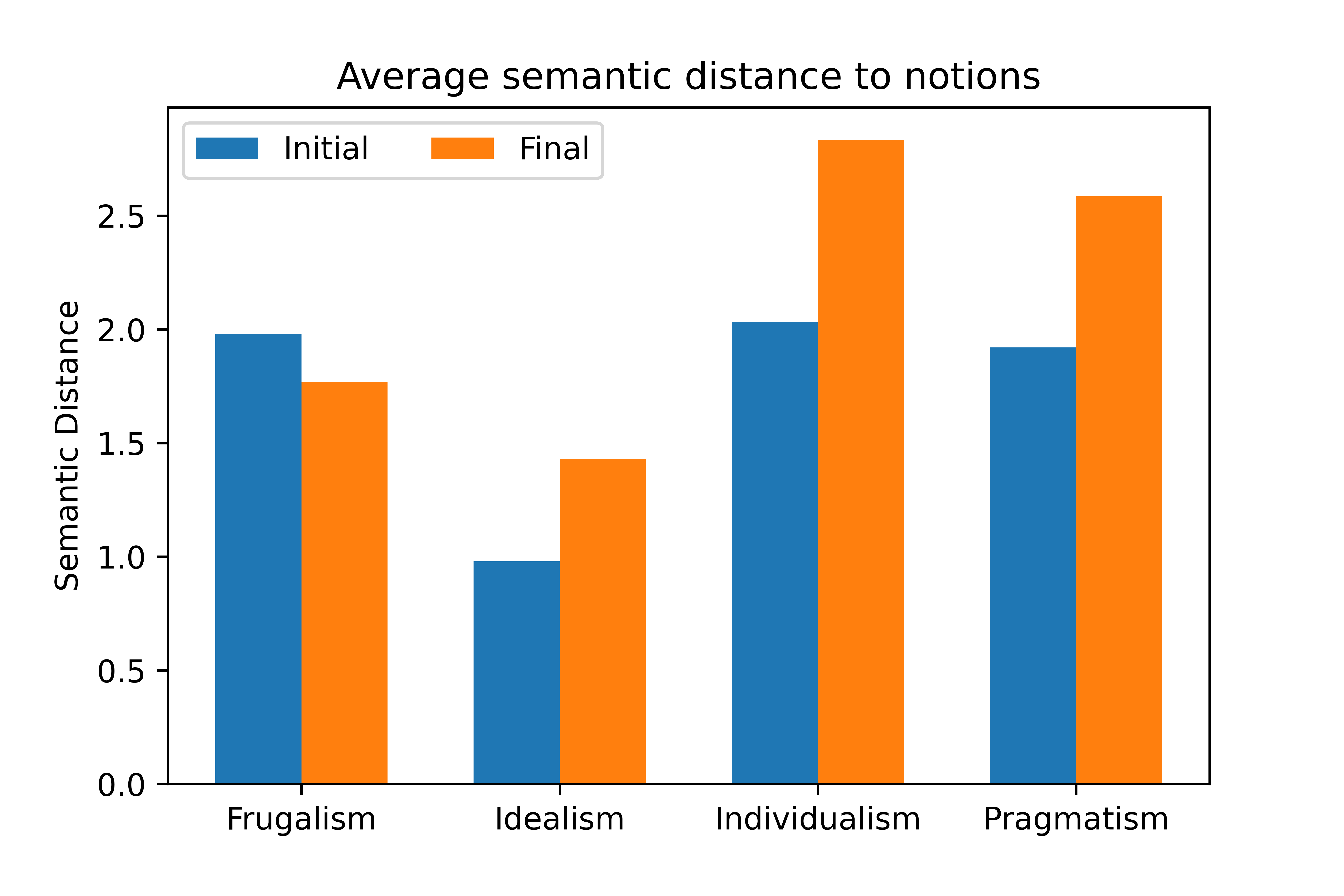} \\
        \hline
       Individualism - Public transit: $21.4\%$ & Pragmatism - Public transit: $33.6\%$\\
       \includegraphics[width=0.4\textwidth]{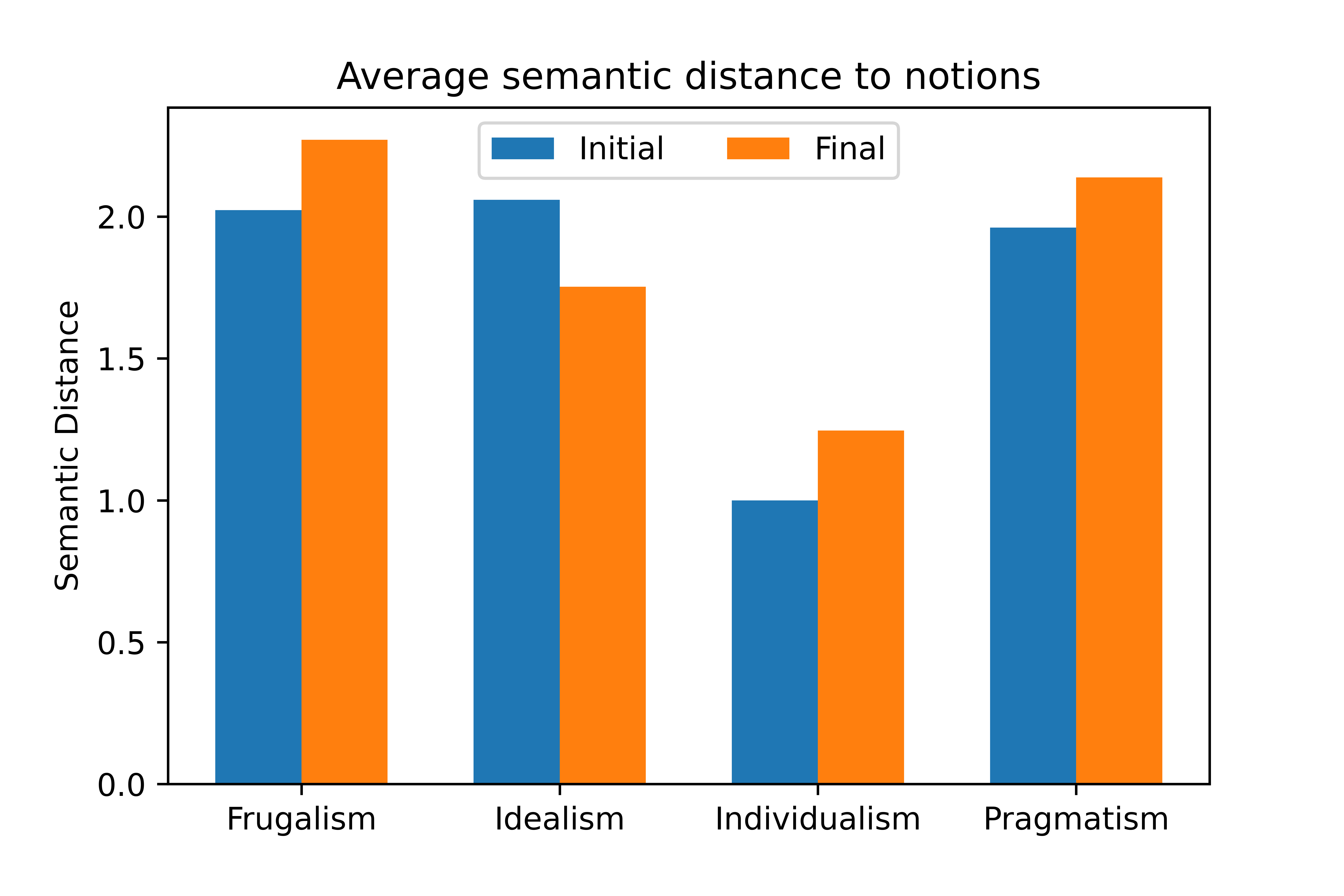} &
       \includegraphics[width=0.4\textwidth]{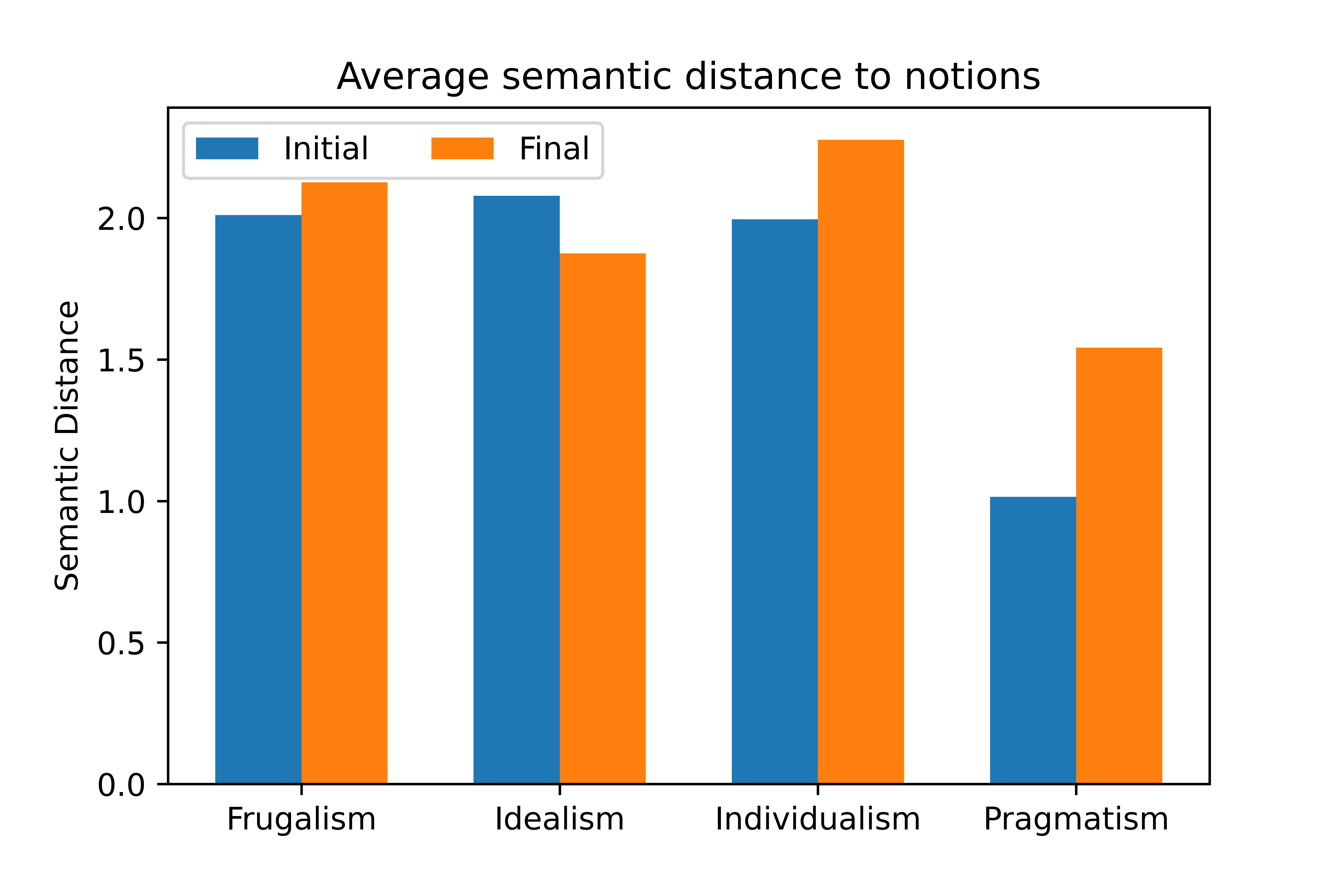}\\
        \hline
    \end{tabular}
    \caption{Initializating reduced average semantic distance to a particular value has lasting effects on the stabilized network.}
    \label{tab:initialDistancesVary}
\end{table}
\vspace{-20pt}
We observe that the initial distributions have a significant impact on the final state of the population, as reflected in the orange bars. This highlights the importance of understanding the population’s prior belief distribution when designing policy interventions aimed at achieving a desirable collective outcome. Additionally, for each configuration in Table~\ref{tab:initialDistancesVary}, we report the average transit choice in the final population. Notably, populations initialized with stronger associations to Frugalism and Individualism exhibit the most extreme behavioral outcomes. This aligns with our earlier findings in Section~\ref{subsec:ImpactValues}, where we identified these values as strong predictors of transit choices.

\subsection{Conformity Leads to Homogeneity}
\begin{figure}[ht]
    \centering
    \vspace{-25pt}
    \includegraphics[width=0.8\columnwidth]{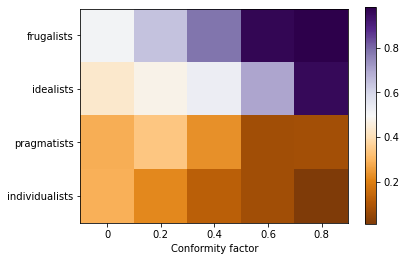}
    \vspace{-10pt}
    \caption{Heatmap of the effect of conformity and initialization of semantic distance to values on the proportion of stabilized population choosing public transport}
    \label{fig:erdosConform}
\end{figure}

The underlying social network structure also induces conformity among agents. We model this by assigning varying levels of the conformity factor ($cf$), as defined in Equation~\ref{eqn:conform}. A higher value of $cf$ indicates a stronger tendency for agents to align their choices with those of their neighbors. Figure~\ref{fig:erdosConform} presents a heatmap illustrating how different levels of conformity, in conjunction with the initial semantic distance configurations from Section~\ref{subsec:initialSD}, influence the proportion of the population choosing public transit. The results show that increased conformity amplifies polarization in transit choices. As $cf$ increases, agents are more likely to adopt the dominant behavior in their local network, leading to stronger collective preferences. Furthermore, populations initially inclined toward Frugalism and Individualism tend to polarize more rapidly compared to those inclined toward Idealism and Pragmatism, highlighting the influence of strong versus weak value indicators of behavior as discussed in Section~\ref{subsec:ImpactValues}.

\vspace{-15pt}

This phenomenon can be interpreted as a shift in the balance between socially derived utility and intrinsic utility based on values. Initially weak inclinations to certain value associations at the population level may, through repeated social reinforcement, become part of an agent's identity, resulting in a more homogenous preference across the population. These findings suggest that external social factors, such as conformity, can significantly reshape internal identity associations. In highly conforming populations, this leads to a reduction in behavioral diversity. Thus, the extended \texttt{CT} framework highlights the emergent role of social influence in shaping collective decision-making dynamics.
    

\section{Conclusion}\label{sec:conclusions}


In this work, we extended the Computational Transcendence (CT) framework to model multi-agent decision-making influenced by identification with abstract human values. Our extended framework enables agents to be adaptive by allowing them to update their identity associations based on interactions with their environment. Through simulation, we observed that agents develop distinct behavioral profiles depending on the values they identify with, and that certain values can exert a stronger influence on specific decisions within a given context. Our findings suggest that the identity-based modeling of values provides a promising direction for designing value-aligned artificial agents. This framework is not only interpretable but also flexible, allowing for integration with domain-specific value encodings.

%
\bibliographystyle{splncs04}
\bibliography{main}

\begin{thebibliography}{10}
\providecommand{\url}[1]{\texttt{#1}}
\providecommand{\urlprefix}{URL }
\providecommand{\doi}[1]{https://doi.org/#1}

\bibitem{deshmukh2022computational}
Deshmukh, J., Srinivasa, S.: Computational transcendence: Responsibility and agency. Frontiers in Robotics and AI  \textbf{9} (2022)

\bibitem{fritzsche1991model}
Fritzsche, D.J.: A model of decision-making incorporating ethical values. Journal of Business Ethics  \textbf{10},  841--852 (1991)

\bibitem{ai_values_alignment}
Gabriel, I.: Artificial intelligence, values, and alignment. Minds \& Machines pp. 411--437 (2020)

\bibitem{hadfield2016}
Hadfield-Menell, D., Russell, S.J., Abbeel, P., Dragan, A.: Cooperative inverse reinforcement learning. Advances in neural information processing systems  \textbf{29} (2016)

\bibitem{kahl2022}
Kahl, S., Wiese, S., Russwinkel, N., Kopp, S.: Towards autonomous artificial agents with an active self: modeling sense of control in situated action. Cognitive Systems Research  \textbf{72},  50--62 (2022)

\bibitem{kai1979prospect}
Kahneman, D., Tversky, A.: Prospect theory: An analysis of decision under risk. Econometrica  \textbf{47}(2),  363--391 (1979)

\bibitem{Kluckhohn_value_action}
Kluckhohn, C.: Values and value-orientations in the theory of action: An exploration in definition and classification. In: Toward a General Theory of Action, pp. 388--433. Harvard University Press, Cambridge, MA and London, England (1951)

\bibitem{railTransitGumbel2016}
Monterola, C., Legara, E.F., Pan, D., Lee, K.K., Hung, G.G.: Non-invasive procedure to probe the route choices of commuters in rail transit systems. Procedia Computer Science  \textbf{80},  2387--2391 (2016)

\bibitem{montes2022synthesis}
Montes, N., Sierra, C.: Synthesis and properties of optimally value-aligned normative systems. Journal of Artificial Intelligence Research  \textbf{74},  1739--1774 (2022)

\bibitem{frameworkHumanValues}
Osman, N., d'Inverno, M.: A computational framework of human values. In: Proceedings of AAMAS 2024. pp. 1531--1539 (2024)

\bibitem{schwartz2012}
Schwartz, S.H.: An overview of the schwartz theory of basic values. Online readings in Psychology and Culture  \textbf{2}(1), ~11 (2012)

\bibitem{value_alignment_formal}
Sierra, C., Osman, N., Noriega, P., Sabater{-}Mir, J., Perell{\'{o}}, A.: Value alignment: a formal approach. CoRR  \textbf{abs/2110.09240} (2021), \url{https://arxiv.org/abs/2110.09240}

\bibitem{Paradigms_of_Computational_Agency}
Srinivasa, S., Deshmukh, J.: Paradigms of computational agency. CoRR  \textbf{abs/2112.05575} (2021), \url{https://arxiv.org/abs/2112.05575}

\bibitem{subagdja2021}
Subagdja, B., Tay, H.Y., Tan, A.H.: Who am i?: Towards social self-awareness for intelligent agents. In: Bessiere, C. (ed.) Proceedings of IJCAI-2020. pp. 4396--4402 (7 2020), special track on AI for CompSust and Human well-being

\end{thebibliography}
\end{document}